# A ferroelectric-like structural transition in a metal

Youguo Shi,[1,2,†] Yanfeng Guo,[1,3,†] Xia Wang,[1] Andrew J. Princep,[3] Dmitry Khalyavin,[4] Pascal Manuel,[4] Yuichi Michiue,[5] Akira Sato,[6] Kenji Tsuda,[7] Shan Yu,[1] Masao Arai,[8] Yuichi Shirako,[9] Masaki Akaogi,[9] Nanlin Wang,[2] Kazunari Yamaura,[1,*] Andrew T. Boothroyd[3,*]

[1] Superconducting Properties Unit, National Institute for Materials Science, 1-1 Namiki, Tsukuba, Ibaraki 305-0044, Japan

[2] Institute of Physics, Chinese Academy of Sciences, Beijing 100190, China

[3] Department of Physics, University of Oxford, Clarendon Laboratory, Oxford, OX1 3PU, U.K.

[4] ISIS Facility, Rutherford Appleton Laboratory, Chilton, Didcot, OX11 0QX, U.K.

[5] Sialon Unit, National Institute for Materials Science, 1-1 Namiki, Tsukuba, Ibaraki 305-0044, Japan

[6] Materials Analysis Station, National Institute for Materials Science, 1-1 Namiki, Tsukuba, Ibaraki 305-0044, Japan

[7] Institute of Multidisciplinary Research for Advanced Materials, Tohoku University, 2-1-1, Katahira, Aoba-ku, Sendai 980-8577, Japan

[8] Computational Materials Science Unit, National Institute for Materials Science, 1-1 Namiki, Tsukuba, Ibaraki 305-0044, Japan

[9] Department of Chemistry, Gakushuin University, 1-5-1 Mejiro, Toshima-ku, Tokyo 171-8588, Japan

† These authors contributed equally to this work.

* E-mail: yamaura.kazunari@nims.go.jp (KY); a.boothroyd@physics.ox.ac.uk (ATB)

**Metals cannot exhibit ferroelectricity because static internal electric fields are screened by conduction electrons[1], but in 1965, Anderson and Blount predicted the possibility of a ferroelectric metal, in which a ferroelectric-like structural transition occurs in the metallic state[2]. Up to now, no clear example of such a material has been identified. Here we report on a centrosymmetric ($R$-$3c$) to non-centrosymmetric ($R3c$) transition in metallic $LiOsO_3$ that is structurally equivalent to the ferroelectric transition of $LiNbO_3$ (ref. 3). The transition involves a continuous shift in the mean position of $Li^+$ ions on cooling below 140K. Its discovery realizes the scenario described in ref. 2, and establishes a new class of materials whose properties may differ from those of normal metals.**

The phase transition considered by Anderson and Blount is a continuous structural transition in the metallic state which must be accompanied by the appearance of a polar axis, and the disappearance of an inversion centre[2]. To the best of our knowledge, transitions satisfying all of these requirements have not been reported. Initially, several A15-type superconductors, including $V_3Si$ and $Nb_3Sn$, were suggested to be ferroelectric metals in the sense of ref. 2, but detailed studies showed the structural transitions to be either weakly first order in the strain or electronically driven[4,5]. In the last decade, a pyrochlore $Cd_2Re_2O_7$ with an unusual structural transition in its metallic state at a temperature of 200 K (refs. 6,7) was suspected to exhibit Anderson and Blount-type metallic ferroelectricity, but a unique polar axis could not be identified and it was concluded that the transition is better described as a piezoelectric transition. Very recently, a reduced titanium oxide $BaTiO_{3-\delta}$ was claimed to display a ferroelectric-like distortion in its metallic state[8]. A neutron diffraction study concluded, however, that the ferroelectric ordering and metallic conduction occur in two distinct phases which do not coexist microscopically[9].

We found that the high-pressure-synthesized material LiOsO$_3$ (see Supplementary Information) shows a structural transition at a temperature $T_s$ = 140 K. The room-temperature crystal structure of LiOsO$_3$ was initially examined using powder X-ray diffraction (XRD). The Goldschmidt diagram predicts that LiOsO$_3$ crystallizes into a LiNbO$_3$-type structure[3,10], and a preliminary refinement of the structure was carried out in the *R*-3*c* space group with Os at the 6*b* site 0,0,0 and O at the 18*e* site *x*,0,1/4. To investigate the position of the Li ion we turned to neutron diffraction, which is much more sensitive to Li than XRD. The neutron diffraction patterns collected above $T_s$ could be successfully described in the *R*-3*c* space group, in agreement with the XRD refinement, with the Li ion at the 6*a* position 0,0,1/4. Atomic absorption spectrometry (see Supplementary Information) indicated that the average Li mass was 2.77%, which corresponds to the composition Li$_{0.98}$OsO$_3$. We have used the stoichiometric composition throughout the structural analysis. The refinement indicated highly anisotropic thermal displacements of the Li ions with considerable extension along the *c*-axis (Table 1 and Fig. 1), which might indicate that the Li ions are distributed equally among equivalent 12*c* sites 0,0,*z* and 0,0,1/2–*z* either side of the oxygen layer at *z* = 1/4, as reported for LiNbO$_3$ and LiTaO$_3$ (refs 3, 11).

The thermal variation of the structure of LiOsO$_3$ was studied by neutron diffraction for temperatures between 10 and 300 K. Figure 1a–d shows structural data obtained from refinements in the *R*-3*c* space group. The lattice parameters (Fig. 1a) decrease uniformly from 300 K until $T_s$ = 140 K, below which the parameter *c* increases and *a* decreases with only a small variation in the unit-cell volume. Just below $T_s$, the non-symmetry-breaking strain components $e_{xx} + e_{yy}$ and $e_{zz}$ vary almost linearly (Fig. 1b). These results show that the phase transition is continuous and the strain components behave like a secondary order parameter coupled to a primary one via a linear–quadratic free energy invariant[12]. The primary order parameter must necessarily be symmetry-breaking according to Landau's theory of second-order phase transitions[12]. Furthermore, the anisotropic thermal parameter $\beta_{33}$, which describes Li displacements along the $c$-axis, increases markedly below $T_s$ (Fig. 1c). This indicates that the primary structural instability involves the position of the Li ions along the c-axis (Fig. 1d).

Given that the phase transition involves a change in symmetry, we find from representation theory[13] that there are three isotropy subgroups, $R$-3, $R$32 and $R3c$, which maintain the translational invariance of the $R$-$3c$ space group and allow the transition to be continuous. These space groups were tested by refinement against the neutron diffraction data at 10 K. Note that $R$-3 and $R$32 should generate additional reflections below $T_s$ which were not observed in the experiment. The refinement in the non-centrosymmetric $R3c$ space group gave the best description of the data (see Table 2 and Fig. 1e). A notable difference between the structures at 10 and 300 K is a shift in the mean position of the Li ion of almost 0.5 Å along the c-axis (Tables 1 and 2).

To confirm the loss of inversion symmetry below $T_s$, we performed convergent-beam electron diffraction (CBED) on crystals of $LiOsO_3$ at room temperature and at 90 K. CBED is able to detect loss of inversion symmetry because, unlike with kinematic diffraction techniques, Friedel's law is not applicable in CBED owing to strong multiple diffraction effects[14]. Details of the method are provided in the Supplementary Information, and the main results are shown in Figs. 2a–d. The projection of the room-temperature CBED pattern taken along the [120] zone axis clearly displays mirror symmetries parallel and perpendicular to the $c^*$-axis (Fig. 2a), in good agreement with a simulation for the $R$-$3c$ structure (Fig. 2c). Thus, the room-temperature structure is unquestionably centrosymmetric. In contrast, the mirror plane perpendicular to $c^*$ is absent at 90 K, both in the data (Fig. 2b) and simulation for the $R3c$ model (Fig. 2d), which confirms that the 90 K structure is non-centrosymmetric. The neutron diffraction and CBED results show conclusively that the inversion center disappears upon cooling through $T_s$ and indicate that the $c$-axis corresponds to a polar axis. The structural features at $T_s$ are completely analogous to those of the isostructural ferroelectrics LiNbO$_3$ and LiTaO$_3$ at their ferroelectric transition temperatures $T_c$ (refs. 3, 11).

The electronic density of states (DOS) of LiOsO$_3$ was calculated by a first principles method using the experimental lattice parameters and atomic coordinates and indicated that the ground state is metallic in nature (Supplementary Fig. S2). In fact, the electrical resistivity ($\rho$) measured on a single crystal showed metallic character over the whole temperature range studied (Fig. 3c). Interestingly, the DOS structures calculated for the $R\bar{3}c$ and $R3c$ models are almost indistinguishable. The electron transport, however, is clearly influenced by the structural phase transition with a marked anomaly in $\rho$ appearing at $T_s$ (Fig. 3c). Above $T_s$, $\rho$ is nearly temperature independent, and this is highly indicative of incoherent charge transport due to disorder scattering of conducting charges[15], which can probably be attributed to the Li disorder in the $R\bar{3}c$ structure. Below $T_s$, a Fermi liquid-like feature appears, suggesting that disorder scattering is reduced by the ordering of Li. The residual resistivity is more than one order of magnitude greater than that expected for a normal metal, implying that defect and domain scattering are substantial, regardless of the Li ordering. Nevertheless, the main features of the $\rho$–$T$ relationship agree well with the centrosymmetric ($R\bar{3}c$) to non-centrosymmetric ($R3c$) transition on cooling.

Structural transitions involving strain are usually discontinuous unless they are driven by a symmetry-breaking primary order parameter[2]. A possible driver for the structural transition could be magnetic ordering. If localized magnetic moments were involved then this could also provide an alternative explanation for the resistivity behaviour. The magnetic susceptibility $\chi$ at $T_s$ is ~$2.7 \times 10^{-4}$ e.m.u. Oe$^{-1}$ mol$^{-1}$, and is only weakly temperature dependent above $T_s$ (Fig. 3b). The calculated electronic DOS at the Fermi energy corresponds to $\chi \approx 9.3 \times 10^{-5}$ e.m.u. Oe$^{-1}$ mol$^{-1}$, indicating that $\chi$ is enhanced compared with the ordinary Pauli paramagnetism. Below $T_s$, Curie–Weiss-like behaviour is observed, indicating more localized paramagnetic character. The statistical quality of the neutron diffraction patterns is such that we would expect to observe magnetic Bragg peaks from any long-range magnetic order with an ordered moment of ~$0.2\mu_B$ or greater. We could not detect any such new Bragg peaks below $T_s$, and neither was there any anomaly at $T_s$ in the integrated background at large $d$-spacing. Taking the susceptibility and neutron diffraction data together, we find no evidence that magnetic order accompanies the structural transition.

The specific heat $C_p$ displays a broad peak near $T_s$ (Fig. 3a), indicating a second-order phase transition. To estimate the transition entropy, $\Delta S$, we fitted a polynomial function using a linear least-squares method to $C_p(T)$ between 20 and 300 K, excluding the transition region (solid curve in Fig. 3a). By subtracting $C_{fit}(T)$ from the data and integrating $(C_p - C_{fit})/T$ we estimate $\Delta S$ to be 0.34$R$ (inset to Fig. 3a), where $R$ is the universal gas constant. This is about half the value $R \ln 2$ expected for a two-site order–disorder transition of the Li ions. The discrepancy could be due to errors in the estimate of the $C_p$ background, or to residual disorder of the Li ions. In the presence of a magnetic field of 70 kOe, the $C_p$ peak changed only slightly, indicating that the entropy is mostly removed by the structural transition. At temperatures below 20 K, a fit to the approximate Debye model $C_p(T)/T = \beta T^2 + \gamma$ yielded $\gamma = 7.7(2)$ mJ mol$^{-1}$ K$^{-2}$ for the Sommerfeld coefficient in zero field, which compares with $\gamma = 6.8$ mJ mol$^{-1}$ K$^{-2}$ from the calculated DOS (Supplementary Fig. S2). In a field of 70 kOe the experimental value for $\gamma$ increases slightly to 9.38(8) mJ mol$^{-1}$ K$^{-2}$, while the value of $\beta$ remains virtually unchanged.

The absence of magnetic features in the transport, neutron diffraction and calorimetric data indicates that LiOsO$_3$ behaves as a normal metal without significant electronic correlations. Hence, the continuous structural transition at 140 K is unlikely to be driven by collective electron dynamics. Moreover, because the octahedrally-coordinated pentavalent Os ion is not Jahn–Teller active, the structural transition is also unlikely to be attributable to a Jahn–Teller distortion. The structural transition of LiOsO$_3$ is very different from that of the isoelectronic compound NaOsO$_3$, which has strong spin–orbit coupling and exhibits a metal–insulator transition accompanied by antiferromagnetic order at 410 K (ref. 16). The properties of LiOsO$_3$ also contrast with those of metallic Cd$_2$Re$_2$O$_7$, which has strong electronic correlations[17,18] and a possible spin polaron state[19] that could be relevant to the piezoelectric-like structural transition at 200 K in Cd$_2$Re$_2$O$_7$ (ref. 6).

On the other hand, our results show that the transition in LiOsO$_3$ is structurally equivalent to those in LiNbO$_3$ and LiTaO$_3$, two technologically important and extensively studied ferroelectrics[3,11,20–23]. This is very surprising because the ferroelectric-like transition in LiOsO$_3$ occurs in the metallic state, and the mechanisms for structural transitions in metals and insulators are not normally the same. The metallic and magnetic properties of LiOsO$_3$ indicate that the transition is unlikely to be caused by any collective electron dynamics, while the structural data provide compelling evidence that the transition mechanism in LiOsO$_3$ is the same as that in other LiNbO$_3$-type ferroelectric materials, most likely driven by a displacive or order–disorder process involving a shift in the mean positions of the Li atoms along the $c$-axis by almost 0.5 Å. As the phase transition in LiOsO$_3$ occurs at an easily accessible temperature it could offer a route to elucidate the still-uncertain phase transition mechanism in LiNbO$_3$ and LiTaO$_3$ whose ferroelectric transition temperatures are much higher ($T_c$ = 1480 K and 940 K, respectively) and therefore difficult to study.

LiOsO$_3$ is thus identified here as an example of a ferroelectric metal, whose existence was first envisaged half a century ago[2] and has now been confirmed. It is possible that materials in this class could display some interesting physical properties. For example, the existence of ferroelectric-like soft phonons could stabilise non-centrosymmetric superconductivity at enhanced temperatures[24].

**References**

1. Lines, M. E. & Glass, A. M. *Principles and Applications of Ferroelectrics and Related Materials* (Oxford University Press, New York, 2001).

2. Anderson, P. W. & Blount, E. I. Symmetry considerations on martensitic transformations: "ferroelectric" metals? *Phys. Rev. Lett.* **14**, 217-219 (1965).

3. Boysen, H. & Altorfer, F. A neutron powder investigation of the high-temperature structure and phase transition in $LiNbO_3$. *Acta Cryst.* **B50**, 405-414 (1994).

4. Testardi, L. R. Structural instability and superconductivity in A-15 compounds. *Rev. Mod. Phys.* **47**, 637-648 (1975).

5. Paduani, C. & Kuhnen, C. A. Martensitic phase transition from cubic to tetragonal $V_3Si$: an electronic structure study. *Eur. Phys. J.* **66**, 353-359 (2008).

6. Sergienko, I. A. *et al.* Metallic "Ferroelectricity" in the Pyrochlore $Cd_2Re_2O_7$. *Phys. Rev. Lett.* **92**, 065501 (2004).

7. Tachibana, M., Taira, N., Kawaji, H. & Takayama-Muromachi, E. Thermal properties of $Cd_2Re_2O_7$ and $Cd_2Nb_2O_7$ at the structural phase transitions. *Phys. Rev. B* **82**, 054108 (2010).

8. Kolodiazhnyi, T., Tachibana, M., Kawaji, H., Hwang, J. & Takayama-Muromachi, E. Persistence of ferroelectricity in $BaTiO_3$ through the insulator-metal transition. *Phys. Rev. Lett.* **104**, 147602 (2010).

9. Jeong, I.-K. *et al.* Structural evolution across the insulator-metal transition in oxygen-deficient $BaTiO_{3-\delta}$ studied using neutron total scattering and Rietveld analysis. *Phys. Rev. B* **84**, 064125 (2011).

10. Navrotsky, A. Energetics and crystal chemical systematics among ilmenite, lithium niobate, and

perovskite structures. *Chem. Mater*. **10**, 2787–2793 (1998).

11. Abrahams, S. C., Buehler, E., Hamilton, W. C., & Laplaca, S. J. Ferroelectric lithium tantalate—III. Temperature dependence of the structure in the ferroelectric phase and the para-electric structure at 940°K. *J. Phys. Chem. Solids*, **34**, 521-532 (1973).

12. Toledano, J. C. & Toledano, P. *The Landau Theory of Phase Transitions* (World Scientific, Singapore, 1987).

13. Campbell, B. J., Stokes, H. T., Tanner, D. E. & Hatch, D. M. ISODISPLACE: a web-based tool for exploring structural distortions. *J. Appl. Cryst.* **39**, 607–614 (2006).

14. Tanaka, M. & Tsuda, K. Convergent-beam electron diffraction. *J. Electron Microsc*. **60**, S245–S267 (2011).

15. Ohgushi, K. *et al.* Structural and electronic properties of pyrochlore-type $A_2Re_2O_7$ ($A$ = Ca, Cd, and Pb). *Phys. Rev. B* **83**, 125103 (2011).

16. Shi, Y. G. *et al.* Continuous metal-insulator transition of the antiferromagnetic perovskite $NaOsO_3$. *Phys. Rev. B* **80**, 161104(R) (2009).

17. Jin, R. *et al.* Fluctuation effects on the physical properties of $Cd_2Re_2O_7$ near 200 K. *J. Phys.: Condens. Matter* **14**, L117–L123 (2002).

18. Hiroi, Z., Hanawa, M., Muraoka, Y. & Harima, H. Correlations and semimetallic behaviors in pyrochlore oxide $Cd_2Re_2O_7$. *J. Phys. Soc. Jpn.* **72**, 21–24 (2003).

19. Storchak, V. G. *et al.* Spin Polarons in the Correlated Metallic Pyrochlore $Cd_2Re_2O_7$. *Phys. Rev. Lett*. **105**, 076402 (2010).

20. Abrahams, S. C., Levinstein, H. J. & Reddy, J. M. Ferroelectric lithium niobate. 5. Polycrystal X-ray diffraction study between 24° and 1200°C. *J. Phys. Chem. Solids* **27**, 1019–1026 (1966).

21. Weis, R. S. & Gaylord, T. K. Lithium niobate: Summary of physical properties and crystal structure, *Appl. Phys.* **A37**, 191–203 (1985).

22. Prokhorov, A. M. & Kuz'minov, Y. S. *Physics and Chemistry of Crystalline Lithium Niobate* (Adam Hilger, Bristol, 1990).

23. Ohkubo, Y. *et al.* Mechanism of the ferroelectric phase transition in $LiNbO_3$ and $LiTaO_3$. *Phys. Rev. B* **65**, 052107 (2002).

24. Yildirim, T. Ferroelectric soft phonons, charge density wave instability, and strong electron-phonon coupling in $BiS_2$ layered superconductors: A first-principles study. *Phys. Rev. B* **87**, 020506(R) (2013).

**Acknowledgments:**

We thank M. Miyakawa (NIMS) for the high-pressure synthesis experiment, H. X. Yang (CAS) and J. Q. Li (CAS) for the ED study, S. Takenouchi (NIMS) for the AAS, and A. Aimi (Gakushuin Univ.), M. Tachibana (NIMS), and M. Terauchi (Tohoku Univ.) for discussion and suggestions. This research was supported in part by the World Premier International Research Center from MEXT, Japan; a Grant-in-Aid for Scientific Research (22246083 and 25289233) from JSPS, Japan; the Funding Program for World-Leading Innovative R&D on Science and Technology (FIRST Program) from JSPS, Japan; the Advanced Low Carbon Technology Research and Development Program (ALCA) of the Japan Science and Technology Agency (JST), Japan; the 973 project of the Ministry of Science and Technology of China (No. 2011CB921701 and 2011CBA00110), China; and the United Kingdom Engineering and Physical Sciences Research Council (EPSRC), grant no. EP/J017124/1.

## Author Contributions

Y.S. and K.Y. conceived the experiments. Y.S. grew and characterized the crystals together with Y.G., X.W., S.Y. and N.W. Y.M. and A.S. conducted crystal structure analysis by XRD. A P., Y.G., D.K. and P.M. performed powder neutron diffraction measurements and analysis. K.T. performed the CBED experiments, data analysis, and data interpretation. M. Arai investigated the electronic structure by first-principles calculation. Y.S. and M. Akaogi investigated the crystal structure stability under high-pressure conditions. K.Y. and A.T.B. supervised the project and co-wrote the paper. All authors discussed the results and reviewed the manuscript.

## Additional information

Supplementary information is available in the online version of the paper (also below). Reprints and permissions information is available online at www.nature.com/reprints. Correspondence and requests for materials should be addressed to K.Y or A.T.B.

## Competing financial interests

The authors declare no competing financial interests.

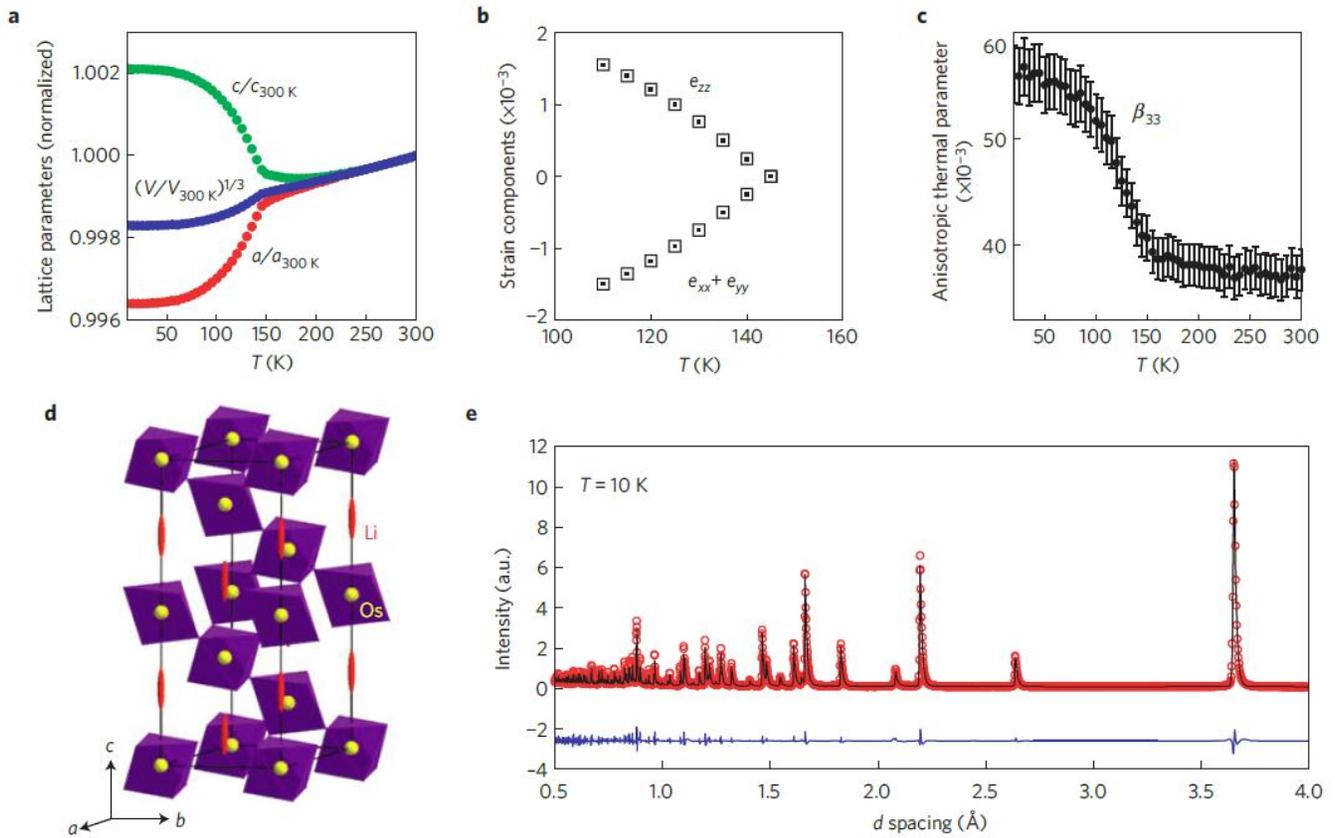

**Figure 1 | Temperature variation of the structural properties of LiOsO$_3$ from Rietveld analysis of neutron diffraction data. a,** Temperature dependence of the lattice parameters through the structural transition at $T_s$ = 140 K. The lattice parameters and unit cell volume are normalized to their values at 300 K. **b,** Strain components $e_{xx} + e_{yy}$ and $e_{zz}$ for temperatures just below $T_s$. The error bars are derived from the statistical errors calculated by the Rietveld refinement program FullProf. **c,** Temperature dependence of the anisotropic thermal parameter $\beta_{33}$, which describes Li displacements along the $c$-axis. **d,** High-temperature centrosymmetric crystal structure of LiOsO$_3$ showing the anisotropic Li displacements (predominantly along the $c$-axis) as 50% probability ellipsoids. **e,** Observed (red circles) and calculated (black line) neutron powder diffraction intensities of LiOsO$_3$ at 10 K. The blue trace below the data shows the difference between the observed and calculated profiles. The refinements were carried out in the $R$-3c space group in **a–d**, and in $R3c$ in **e**.

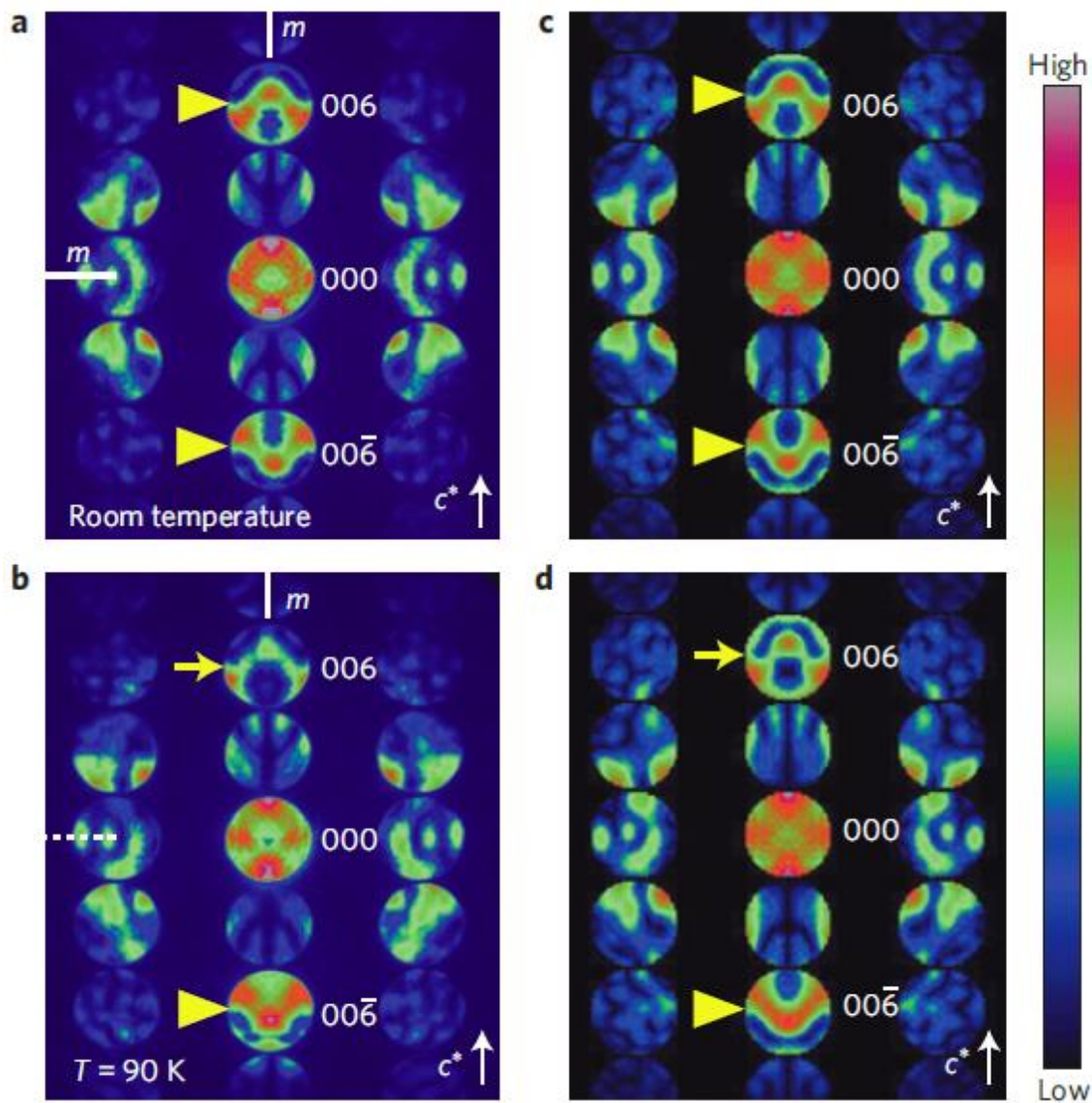

**Figure 2 | Experimental and simulated CBED patterns for LiOsO$_3$ taken along the [120] zone axis.**

**a, b,** Measurements made at room temperature (**a**), and 90 K (**b**). **c, d,** Corresponding simulated CBED patterns for a specimen thickness of 73 nm using the centrosymmetric model (*R-3c*) (**c**) and the non-centrosymmetric model (*R3c*) (**d**). An arrow or arrowhead indicates the absence or presence of mirror symmetry perpendicular to the *c\** axis.

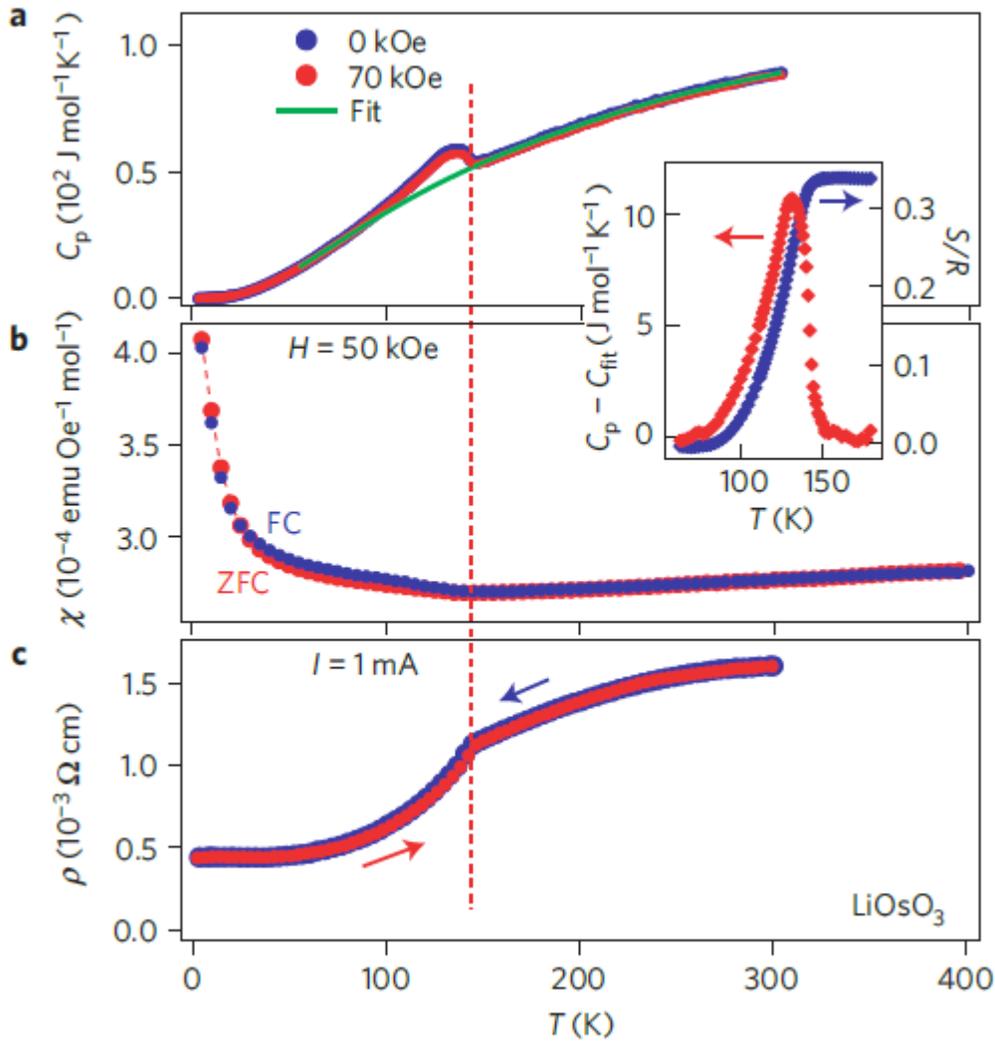

**Figure 3 | Temperature variation of the electrical, magnetic and calorimetric properties of LiOsO$_3$.**

**a**, The heat capacity $C_p$ of a polycrystalline pellet (of mass 7.5 mg). The green line is a fit made to the data outside the transition region. The inset shows the heat capacity in the transition region corrected for the fitted background together with an estimate of the transition entropy obtained by integrating $(C_p - C_{fit})/T$. **b**, The zero-field cooled (ZFC) and field-cooled (FC) magnetic susceptibility $\chi$ of a powder sample (of mass 138 mg) in a measuring field of 50 kOe. **c**, The resistivity $\rho$ measured on a single crystal. Red and blue symbols and arrows indicare data recorded while heating and cooling, respectively. The red vertical dashed line indicates $T_s$.

**Table 1. Structural parameters of LiOsO$_3$ refined from neutron diffraction data collected at 300K**

| Atom | Wyckoff site | x | y | z | $B_{iso}$ (Å$^2$) | $\beta_{ij}$ |
|---|---|---|---|---|---|---|
| Li | 6a | 0 | 0 | 0.25 | - | $\beta_{11}=\beta_{22}=2\beta_{12}=0.007(4)$ |
|    |    |   |   |      |   | $\beta_{33}=0.039(2)$ |
| Os | 6b | 0 | 0 | 0 | 0.93(4) | - |
| O  | 18e | 0.6298(2) | 0 | 0.25 | 1.2(4) | - |

Space group *R*-3*c* (No. 167); cell parameters *a* = 5.06379(5) Å, *c* = 13.2110(2) Å; *V* = 293.371(6) Å$^3$; $R_{Bragg}$ = 4.42%; thermal displacements of the Os and O atoms are described with an isotropic displacement parameter $B_{iso}$, whereas the Li displacements are refined with anisotropic displacement parameters $\beta_{ij}$. Numbers in parentheses are fitting errors in the last digit.

**Table 2. Structural parameters of LiOsO$_3$ refined from neutron diffraction data collected at 10 K**

| Atom | Wyckoff site | x | y | z | $B_{iso}$ (Å$^2$) |
| --- | --- | --- | --- | --- | --- |
| Li | 6a | 0 | 0 | 0.2147(6) | 1.9(2) |
| Os | 6a | 0 | 0 | 0 | 0.79(4) |
| O | 18b | 0.6260(5) | −0.0102(9) | 0.2525(6) | 0.96(4) |

Space group, $R3c$ (No. 161); cell parameters $a$ = 5.04556(5) Å, $c$ = 13.2390(2) Å; $V$ = 291.880(6) Å$^3$; $R_{Bragg}$ = 4.51%; isotropic displacement parameters are used for all atoms. Numbers in parentheses are fitting errors in the last digit.

Supplementary Information for:

# A ferroelectric-like structural transition in a metal

Youguo Shi,[1,2,†] Yanfeng Guo,[1,3,†] Xia Wang,[1] Andrew Princep,[3] Dmitry Khalyavin,[4] Pascal Manuel,[4] Yuichi Michiue,[5] Akira Sato,[6] Kenji Tsuda,[7] Shan Yu,[1] Masao Arai,[8] Yuichi Shirako,[9] Masaki Akaogi,[9] Nanlin Wang,[2] Kazunari Yamaura,[1,*] Andrew T. Boothroyd[3,*]

[1] Superconducting Properties Unit, National Institute for Materials Science, 1-1 Namiki, Tsukuba, Ibaraki 305-0044, Japan

[2] Institute of Physics, Chinese Academy of Sciences, Beijing 100190, China

[3] Department of Physics, University of Oxford, Clarendon Laboratory, Oxford, OX1 3PU, U.K.

[4] ISIS Facility, Rutherford Appleton Laboratory, Chilton, Didcot, OX11 0QX, U.K.

[5] Sialon Unit, National Institute for Materials Science, 1-1 Namiki, Tsukuba, Ibaraki 305-0044, Japan

[6] Materials Analysis Station, National Institute for Materials Science, 1-1 Namiki, Tsukuba, Ibaraki 305-0044, Japan

[7] Institute of Multidisciplinary Research for Advanced Materials, Tohoku University, 2-1-1, Katahira, Aoba-ku, Sendai 980-8577, Japan

[8] Computational Materials Science Unit, National Institute for Materials Science, 1-1 Namiki, Tsukuba, Ibaraki 305-0044, Japan

[9] Department of Chemistry, Gakushuin University, 1-5-1 Mejiro, Toshima-ku, Tokyo 171-8588, Japan

† These authors contributed equally to this work.

* Email: yamaura.kazunari@nims.go.jp (KY); a.boothroyd@physics.ox.ac.uk (ATB)

## 1. Preparation and characterization of materials

Polycrystalline and single-crystal LiOsO$_3$ were prepared by solid-state reaction under high pressure from a mixture of Li$_2$O (97%, Aldrich), OsO$_2$ (Os-84.0%, Alfa Aesar), and LiClO$_4$ (99.99%, Aldrich) in the molar ratio Li$_2$O/OsO$_2$/LiClO$_4$ = 0.5:1:0.125. The mixture was placed in a platinum capsule (6.9 mm in diameter and ~5 mm in height) and maintained at a pressure of 6 GPa in a belt-type high-pressure apparatus and at a temperature of 1200 °C for 1 hour (polycrystals) or 1600 °C for 5 hours (single crystals). LiClO$_4$ probably decomposed to LiCl and O$_2$ at the elevated temperature; LiCl may help the growth of crystals as a flux under these conditions, as well as the small temperature gradient ($\Delta T <$ ~50 K) in the capsule. After heating, the capsule was quenched to room temperature within a few seconds before the pressure was released.

We investigated the crystal structure of a single crystal of LiOsO$_3$ (inset, Fig. 1) by XRD. The crystal was mounted on the edge of a fine glass fiber, which was fixed in a diffractometer equipped with an area detector (SMART APEX, Bruker). The analysis failed, however, because of complex diffraction spots, which were likely caused by domain structures rather than the lattice. More than 10 attempts with different crystals were made, but a completely satisfactory result was not obtained.

Instead, we conducted a Rietveld analysis on the powder XRD pattern obtained from polycrystalline LiOsO$_3$ (Fig. S1 and Table S1). We used monochromatic Cu-K$\alpha$ radiation ($\lambda$ = 1.5418 Å) in a 2$\theta$ range from 20º to 130º, and the software JANA2006 [1] to analyze the pattern. The March-Dollase formula was used to correct for preferred-orientation effects.[2] Crystal structure visualization was achieved using the software VESTA.[3]

In order to measure the net Li content, the polycrystalline LiOsO$_3$ was washed well in water and studied in an atomic absorption spectrometer (Varian Spectra, AA20). We used a Li standard solution (Kanto Chemical Co.) as a reference. The average Li content was 2.77%, which corresponds to the composition Li$_{0.98}$OsO$_3$.

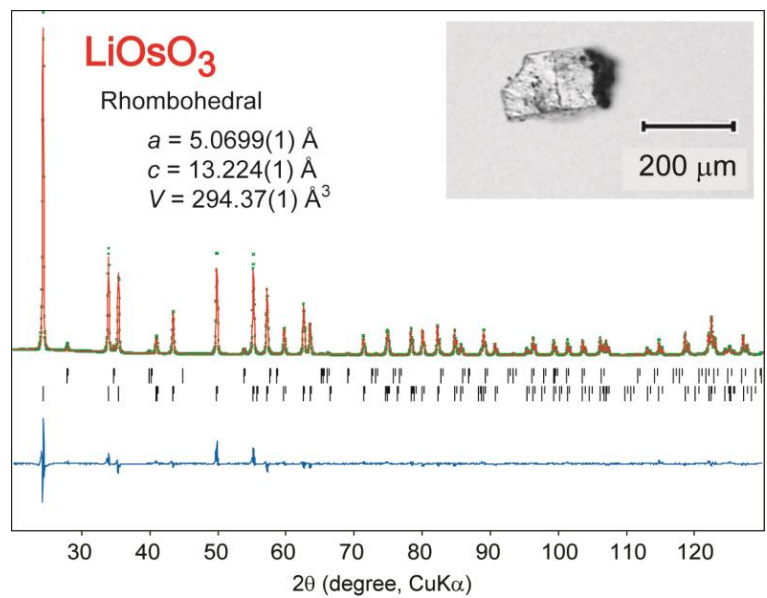

**Figure S1** Rietveld analysis of the room temperature XRD profile of $LiOsO_3$. Small crosses and the solid curve show the observed and calculated profiles, respectively. The difference between the two is shown in blue near the bottom. Vertical ticks indicate Bragg reflections expected for $LiOsO_3$ (lower) and $OsO_2$ (upper). (Inset) Photograph of a selected flux-grown $LiOsO_3$.

We measured the electrical resistivity of a single crystal of $LiOsO_3$ between temperatures of 2 K and 300 K using a four-probe technique with a gauge current of 1 mA on warming and cooling. DC-magnetic susceptibility of the polycrystalline $LiOsO_3$ (138 mg) was measured using a commercial instrument (Magnetic Properties Measurement System, Quantum Design) between 2 K and 400 K with an applied magnetic field of 50 kOe. The powder was loosely gathered in a sample holder and cooled to 2 K. The magnetic field was then applied to the sample, followed by warming to 400 K (zero-field cooling) and cooling to 2 K (field cooling).

The temperature and magnetic-field dependence of $C_p$ of $LiOsO_3$ were measured at temperatures between 3 K and 300 K by a quasi-adiabatic method in a commercial instrument (Physical Properties Measurements System, Quantum Design). A dense pellet of polycrystalline $LiOsO_3$ (7.5 mg) was prepared at 6 GPa in a high-pressure apparatus without heating.

**Table S1**     Interatomic distances and bond angles in LiOsO$_3$.

| Bond | Bond length (Å) | Bond | Bond angle (degree) |
|---|---|---|---|
| Os-O$^i$×2 | 1.9181(7) | Os-O$^i$-Os | 145.5(4) |
| Os-O$^{ii}$×2 | 1.918(2) | O$^i$-Os-O$^{ii}$ | 89.72(12) |
| Os-O$^{iii}$×2 | 1.918(6) | O$^i$-Os-O$^{iii}$ | 89.7(2) |
| Average | 1.918 | O$^{ii}$-Os-O$^{iii}$ | 90.3(2) |
| Effective coordination number $^a$: 6.00 | | | |
| BVS $^b$ | 5.24 | | |
| | | | |
| Li-O$^{iv}$ | 1.988(4) | O$^{iv}$-Li-O$^v$ | 117.9(3) |
| Li-O$^v$ | 1.988(4) | O$^{iv}$-Li-O$^{vi}$ | 117.9(4) |
| Li-O$^{vi}$ | 1.988(9) | O$^v$-Li-O$^{vi}$ | 117.9(4) |
| Li-O$^{vii}$ | 2.47(2) | | |
| Li-O$^{viii}$ | 2.47(2) | | |
| Li-O$^{ix}$ | 2.47(2) | | |
| Average | 2.23 | | |
| Effective coordination number: 3.54 | | | |
| BVS | 0.928 (Li) | | |
| Li-Li $^c$ | 0.58(4) | | |

Symmetry codes: (i) $x, y, z$; (ii) $y, -x + y, -z$; (iii) $x - y, x, -z$; (iv) $x - y + 2/3, x + 1/3, -z + 1/3$; (v) $y + 2/3, -x + y + 1/3, -z + 1/3$; (vi) $-x + 2/3, -y + 1/3, -z + 1/3$; (vii) $x - y, -y, -z + 1/2$; (viii) $-x, -x + y, -z + 1/2$; (ix) $y, x, -z + 1/2$

$^a$ Ref. 5
$^b$ Bond valence sums.
$^c$ Intersite distance

## 2.     Neutron diffraction

Neutron diffraction data were collected on the WISH time-of-flight diffractometer[5] at the ISIS Facility of the Rutherford Appleton Laboratory, UK. A powder sample of mass 4 g was loaded into a 6 mm diameter vanadium cylinder mounted on a closed-cycle refrigerator, and measured on warming between 10 K and 300 K in 5 K steps. The Rietveld refinement of the crystal structure was performed using FullProf[6] against data measured in detector banks at average 2θ values of 58°, 90°, 122° and 154°, each covering 32° of the scattering plane. There were no detectable impurity peaks in the patterns.

## 3. Convergent beam electron diffraction (CBED)

The CBED method has been established as the most powerful technique for determining crystal symmetries of specimens with areas of a few nanometers in diameter owing to the nanometer-size electron probe employed and the strong dynamical diffraction (multiple diffraction) effect.[7–9] In particular, it is frequently used to analyze the electrostatic potential in ferroelectric materials. CBED experiments were performed using a transmission electron microscope (JEM-2010FEF) equipped with an energy-filter, which is very effective for reducing the background intensities of inelastically scattered electrons. CBED patterns were obtained from defect-free single domain areas of a few nm in diameter at an accelerating voltage of 100 kV with an acceptance energy width of approximately 10 eV.

Simulations of CBED patterns were performed on the basis of the many-beam Bloch-wave dynamical diffraction theory using the software MBFIT developed by Tsuda and Tanaka.[10,11] Both zeroth-order Laue zone (ZOLZ) and higher-order Laue zone (HOLZ) reflections were accounted for in the simulations. The direction of polarization can be determined by comparing experimental patterns with simulated ones.

All possible symmetries appearing in the CBED patterns were tabulated by Buxton *et al*. for all 32 point groups.[7] CBED patterns of the ZOLZ reflections, as shown in Figs. 2a and 2b of the main article, contain broad patterns due to ZOLZ reflections and fine lines caused by interactions with the HOLZ reflections. The broad patterns exhibit the symmetry elements of the specimen projected along the zone axis, which are called projection symmetries and could be higher symmetries than those of the patterns including fine lines. In the [120] CBED patterns shown in Figs. 2a and 2b, it is deduced from the Buxton's table that the projection symmetry is expected to be *2mm* for the centrosymmetric space group of *R*-3*c*, and *m* (parallel to the $c^*$-axis) for the non-centrosymmetric space group of *R*3*c*. Thus, these two space groups can be distinguished from the projection symmetry. It is noted that the full symmetry, including fine-line patterns, is *m* (parallel to the $c^*$-axis) for both the *R*-3*c* and *R*3*c* space groups.

## 4. Electronic structure

The electronic structure of LiOsO$_3$ was calculated within the generalized gradient approximation[12] of density functional theory. We used the WIEN2k program[13], which is based on the full-potential augmented plane-wave (APW) methods. The spin-orbit interaction is included as a perturbation to the scalar-relativistic equations. The cut-off wave vector K for APW basis sets was chosen as RK = 8, where R is smallest muffin-tin radius, *i.e.*, 1.6 (a.u.) for oxygen atoms. The integration over the Brillouin zone was approximated by a tetrahedron method with 182 k-points in an irreducible zone.

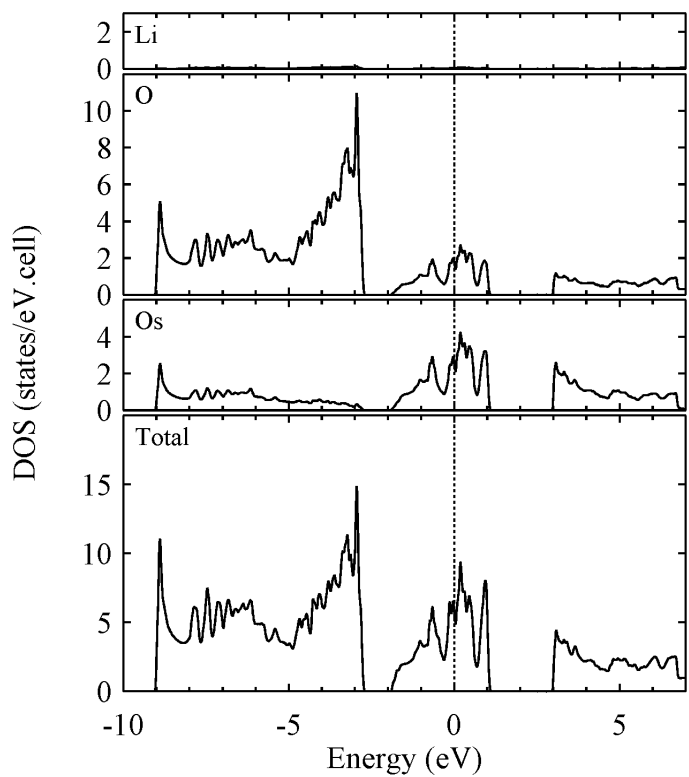

**Figure S2** **Total and partial densities of states (DOS) for LiOsO$_3$.** The vertical dotted line indicates the Fermi energy.

**Supplementary references**

1. Petricek, V., Dusek, M. & Palatinus, L. *Jana2006: A crystallographic Computing System* (Institute of Physics, Praha, Czech Republic, 2006).

2. Dollase, W. A. Correction of intensities for preferred orientation in powder diffractometry: application of the March model. *J. Appl. Cryst.* **19**, 267–272 (1986).

3. Momma, K. & Izumi, F. *VESTA 3* for three-dimensional visualization of crystal, volumetric and morphology data. *J. Appl. Cryst.* **44**, 1272–1276 (2011).

4. Nespolo, M., Ferraris, M. & Hoppe R. Charge distribution analysis of ceramic materials. *J. Ceram. Process. Res.* **2**, 38–41 (2001).

5. Chapon, L. C. *et al.* Wish: the new powder and single crystal magnetic diffractometer on the second target station. *Neutron News* **22**, 22–25 (2011).

6. Rodriguez–Carvajal, J. Recent advances in magnetic structure determination by neutron powder diffraction. *Physica B* **192**, 55–69 (1993); http://www.ill.eu/sites/fullprof/.

7. Buxton, B. F., Eades, J. A., Steeds, J. W. & Rackham, G. M. The symmetry of electron diffraction zone axis patterns. *Phil. Trans. R. Soc. London* **281**, 171–194 (1976).

8. M. Tanaka, *International Tables for Crystallography*, 3rd edn., Vol. B, pp 307–356, edited by U. Shumueli, IUCr, Springer, Dordrecht (2008).

9. Tanaka, M. &Tsuda, K. Convergent-beam electron diffraction. *J. Electron Microsc.* **60**, S245–67 (2011).

10. Tsuda. K. and Tanaka, M. Refinement of crystal structure parameters using two-dimensional energy-filtered CBED patterns. *Acta Cryst. A* **55**, 939–954 (1999).

11. Tsuda, K. Ogata, Y., Takagi, K., Hashimoto, T. & Tanaka, M. Refinement of crystal structural parameters and charge density using convergent-beam electron diffraction — the rhombohedral phase of LaCrO$_3$. *Acta Cryst. A* **58**, 514–525 (2002).

12. Perdew, J. P., Burke, K. & Ernzerhof, M. Generalized gradient approximation made simple, *Phys. Rev. Lett.* **77**, 3865–3868 (1996).

13. Blaha, P. Schwarz, K., Madsen, G. K. H., Kvasnicka, D. & Luitz, J. *WIEN2k, An Augmented Plane Wave + Local Orbitals Program for Calculating Crystal Properties*; K. Schwarz, Ed. (Tech. Univ. Wien, Austria, 2001).